# Robust antiferromagnetism preventing superconductivity in pressurized $Ba_{0.61}K_{0.39}Mn_2Bi_2$

Dachun Gu[1], Xia Dai[1], Congcong Le[1], Liling Sun[1,2]†, Qi Wu[1], Bayrammurad Saparov[3], Jing Guo[1], Peiwen Gao[1], Shan Zhang[1], Yazhou Zhou[1], Chao Zhang[1], Lun Xiong[4], Rui Li[4], Yanchun Li[4], Xiaodong Li[4], Jing Liu[4], Athena S. Sefat[3], Jiangping Hu[1,2] and Zhongxian Zhao[1,2]†

[1]Institute of Physics and Beijing National Laboratory for Condensed Matter Physics, Chinese Academy of Sciences, Beijing 100190, China
[2]Collaborative Innovation Center of Quantum Matter, Beijing, 100190, China
[3]Materials Science and Technology Division, Oak Ridge National Laboratory, Oak Ridge, TN 73831-6056, USA
[4]Institute of High Energy Physics, Chinese Academy of Sciences, Beijing 100049, China

$BaMn_2Bi_2$ possesses an isostructure of iron pnictide superconductors with $ThCr_2Si_2$-type structure and an antiferromagnetic (AFM) ground state similar to that of cuprates. It becomes an AFM metal when doped with potassium, therefore, much attention is paid to the properties of such a doped compound when its AFM order is eliminated by pressure, with the expectation of discovering new family of superconductors and understanding the mechanism of high temperature of superconductors. Here, we report that K-doped $BaMn_2Bi_2$ shows no experimental evidence of superconductivity down to 1.5 K under pressures up to 35.6 GPa, however, a tetragonal (T) to an orthorhombic (OR) phase transition is observed at pressure of ~20 GPa. Theoretical calculations for the T and OR phases, on basis of our high-pressure XRD data, find that the AFM order is robust in both of the phases in pressurized $Ba_{0.61}K_{0.39}Mn_2Bi_2$. Our experimental and theoretical results suggest that the K-doped $BaMn_2Bi_2$ belongs to a strong Hund's AFM metal with a hybridization of localized spin electrons and itinerant electrons, and that its robust AFM order essentially prevents the emergence of superconductivity.

PACS number(s): 74.70.Xa, 74.62.Fj

Superconductivity in unconventional high-temperature (high-$Tc$) superconductors is related with an antiferromagnetic (AFM) ground state in a layered undoped 'parent' compound [1-5]. Consequently, layered AFM compounds with higher Neel temperature ($T_N$) are believed to be good candidates for the parent compound of new high-$Tc$ superconductors [6]. The layered Mn-based compounds with the same $ThCr_2Si_2$-type structure (122) as $BaFe_2As_2$ [7], such as $BaMn_2P_2$ [8-10], $BaMn_2As_2$ [11-19] and $BaMn_2Bi_2$ [20-22], have AFM ground state similar to those of a few parent compounds of cuprates [23]. For such similarities, it is reasonably expected that Mn-based 122s might give superconductivity and may potentially bridge the gap between the cuprates and FeAs-based superconductors.

Previous studies on the parents compounds of FeAs-based superconductors showed that both chemical doping and external pressure can efficiently suppress the AFM order and drive the samples into a superconducting state [2,4,24-35]. Recent investigations on $BaMn_2As_2$ found that chemical doping has no significant effect on suppressing AFM order [15-17]. $BaMn_2Bi_2$ is an AFM semiconductor at ambient pressure, the properties of which lie in the intermediary region between two parent compounds of FeAs-122 and $La_2CuO_4$ superconductors. Neutron studies show that $BaMn_2Bi_2$ adopts G-type AFM structure with $T_N = 387$ K in the tetragonal phase [21], alternative from the parent compounds of cuprates and FeAs-122 superconductors. Since potassium (K) doping on Ba site has turned the sample from an AFM semiconductor to an AFM metal [20], it is of great interest to explore the potential superconductivity through suppressing its AFM order by pressure. In this work, we perform high-pressure studies on the Mn-based compound $Ba_{0.61}K_{0.39}Mn_2Bi_2$ by *in-situ* high pressure electrical transport and X-ray diffraction (XRD) measurements in a diamond anvil cell (DAC). We find no sign of pressure-induced superconductivity in $Ba_{0.61}K_{0.39}Mn_2Bi_2$ up to 35.6 GPa, however, a pressure-induced tetragonal-to-orthorhombic (T-to-OR) phase transition is observed. Our calculations reveal that the $Ba_{0.61}K_{0.39}Mn_2Bi_2$ has robust antiferromagnetism in both of the T phase and OR phase and is a strong Hund's AFM metal with a hybridization of localized spin electrons and itinerant electrons.

High quality single crystals with nominal composition $Ba_{0.4}K_{0.6}Mn_2Bi_2$ were synthesized by the similar method as described in Ref. [20]. High-pressure resistance measurements using the standard four-probe method were performed in a DAC made from Be-Cu alloy in a house-built refrigerator. Diamond anvils of 500 μm and 300 μm flats and rhenium gaskets with 200 μm and 100 μm diameter sample holes were used, respectively, for different runs. NaCl powders were employed as pressure medium for the high-pressure resistance measurements. Structural information under pressure was obtained through the angle-dispersive powder XRD experiments, performed on beamline 4W2 at the Beijing Synchrotron Radiation Facility (BSRF). Diamonds with low birefringence were selected for the high-pressure XRD experiments. Diamond anvils of 400 μm flat and a stainless steel gasket with a 150 μm diameter sample hole were used. A monochromatic X-ray beam with a wavelength of 0.6199 Å was adopted for all measurements. The XRD images were collected using Mar345 detector, and the XRD geometry was calibrated by $CeO_2$. To keep the sample in a quasi-hydrostatic pressure environment, silicon oil was used for the XRD measurements. Pressure was determined by ruby fluorescence method [36]. Since the sample is air sensitive, the samples either for high-pressure resistance or XRD measurements were loaded into the DAC in a glove-box.

At ambient pressure, we characterized the quality, actual composition and physical property of the sample investigated. As shown in Fig.1a, sharp (00l) peaks in the diffraction pattern indicate the high quality of the resulting sample. To determine the actual composition for the K-doped sample, we did energy dispersive X-ray (EDX) analysis using a Hitachi S-4800 scanning electron microscope equipped with a HORIBA EMAX EDX analysis system (Fig. 1b and 1c). The actual doping level of K in the sample is about 0.39. Electrical resistance measurement shows that the $Ba_{0.61}K_{0.39}Mn_2Bi_2$ is a metal at ambient pressure (Fig 1d), consistent with literature [20].

Figure 2(a) and 2(b) show temperature ($T$) dependent of resistance ($R$) of the $Ba_{0.61}K_{0.39}Mn_2Bi_2$ under pressure up 35.6 GPa. These data are remarkable in that the *R-T* curves are pressure dependent, *i.e.* the *R-T* curve moves up with increasing pressure up to 19.9 GPa, while it changes its trend for applied pressure ranging from 19.9 to 35.6 GPa. Plot of pressure dependence of resistance measured at different temperatures illustrates this conspicuous feature (Fig.2c). This feature may be associated with a structure phase transition. Zooming in the *R-T* curves for lower temperature range, we found no sign of pressure-induced superconductivity down to 1.5 K under pressure up to 35.6 GPa (Fig.2d and Fig.2e). To confirm the experimental results obtained, we loaded the single crystal with the actual composition of Ba$_{0.68}$K$_{0.32}$Mn$_2$Bi$_2$ from Sefat's Group into a DAC in a glove-box and performed experiments in the same manner. We observed the same high pressure behavior as that of Ba$_{0.61}$K$_{0.39}$Mn$_2$Bi$_2$.

Structure information is crucial in tailoring the superconductivity. At ambient pressure, BaMn$_2$Bi$_2$ adopts a body-centered ThCr$_2$Si$_2$ tetragonal structure in the space group *I*4/*mmm*. Hole-doping via substitution of Ba with K in the form of Ba$_{1-x}$K$_x$Mn$_2$Bi$_2$ does not alter its crystal structure [20]. Applying pressure on the Ba$_{0.61}$K$_{0.39}$Mn$_2$Bi$_2$ found a T-to-OR phase transition at pressure of 19.8 GPa, as shown in Fig.3(a). The OR phase in Ba$_{0.61}$K$_{0.39}$Mn$_2$Bi$_2$ persists up to 28.8 GPa. The evolution of crystal lattice parameters and volume with pressure in the two phases are displayed in Fig. 3(b) and 3(c).

Figure 4(a) and 4(b) present XRD patterns of the pressurized sample and corresponding Rietveld refinement results. It is seen that the XRD pattern obtained at 1.7 GPa can well be refined in the T phase in the *I*4/*mmm* space group, yielding the reliability factors of $R_p$=3.34% and $R_{wp}$=5.13%, respectively, as well as the fitting goodness $\chi$<1. The refinement of the XRD data collected at 19.8 GPa is in good agreement with OR phase in the *Fmmm* space group; the reliability factors are $R_p$=1.77%, $R_{wp}$ =2.84% and the $\chi$<1, respectively. Figure 4(c) shows the X-ray diffraction images for the Ba$_{0.61}$K$_{0.39}$Mn$_2$Bi$_2$ at different pressures. It is seen that the degree of crystallinity has got poor with further increasing pressure. At 28.8 GPa, a halo-like ring is observed, together with corresponding broadening XRD pattern at this pressure, suggesting that part of the OR phase has transformed to an amorphous-like phase.

The pressure induced T-to-OR phase transition determined by high-pressure XRD measurements is consistent with our resistance data. As shown in Fig.2c, pressure dependence of resistance displays a dome-like feature. The resistance increases with pressure, reaches a maximum at ~20 GPa and then decreases with further increasing pressure. Notably, the change in resistance against pressure follows the similar trend at different temperatures down to 4 K, indicating that either the room-temperature T phase or the room-temperature OR phase can be maintained down to 4 K.

With pressure-induced volume shrinking, the band position and density-of-states (DOS) may change correspondingly. To identify whether the pressure may induce a significant change in electronic structure, we carried out theoretical calculations on basis of our high-pressure XRD data for the T and OR phases of $Ba_{1-x}K_xMn_2Bi_2$. Our calculations were performed using density functional theory (DFT) as implemented in the Vienna *ab* initio simulation package (VASP) code [37]. The Perdew-Burke-Ernzerhof (PBE) exchange-correlation functional [38] and the projector-augmented-wave (PAW) approach [39] are used. Throughout the work, the cutoff energy is set to be 400 eV. The positions of all the atoms are fully relaxed during the geometry optimizations with forces minimized to less than 0.01 eV/Å. On the basis of the equilibrium structure, 20 k points are used to compute the band structure. We have also performed GGA+U calculations，where U is the onset coulomb repulsive energy of Mn. With a modest U value (2.5 eV), we find that the result reported in the following for both band structures and magnetisms are not qualitatively modified.

Our calculation results are summarized in Fig. 5. We find that the ground state of the pressurized $Ba_{0.61}K_{0.39}Mn_2Bi_2$ is a robust G-type AFM metal in the T phase with an ordered magnetic moment ~3.4$\mu_B$/Mn aligned parallel to the $c$ axis. The band structure and DOS of $Ba_{0.61}K_{0.39}Mn_2Bi_2$ in the paramagnetic state and AFM state at 17.49 GPa are shown in Fig. 5(a) and Fig. 5(b), respectively. Both states are metallic. In the AFM state, the total energy per unit cell which includes 4 Mn atoms is 3.29 eV, lower than that in the paramagnetic state, indicating that the AFM state is the ground state of this compound. In the paramagnetic state, the 3$d$ states of Mn dominate near Fermi level as shown Fig. 5(a). However, in the AFM state, there is large weight redistribution for the 3$d$ states of Mn. The 3$d$ electron density has large concentration from -3 eV to -2 eV and from 0.5 eV to 2 eV, which indicates the large energy splitting caused by the AFM ordering, consistent with the large local spin moment formed through strong Hund's coupling [40]. In the OR phase, the ground state is still a G-type AFM state with a magnetic moment of ~3.4 $\mu_B$/Mn. Moreover, the calculation results on OR phase show that the direction of magnetic moment is parallel to the $a$-axis, which differs from the direction along the $c$-axis in the T phase. This result is consistent with the obvious orthorhombic lattice distortion measured in our high-pressure XRD experiments. As shown in Fig.5(c) and Fig.5(d), the band structure and DOS at 24.6 GPa in both paramagnetic state and AFM states in the OR phase are almost unchanged, comparing with that at 17.49 GPa, indicating that pressure cannot significantly suppress the AFM order in $Ba_{0.61}K_{0.39}Mn_2Bi_2$.

In conclusion, a combination of high pressure resistance and XRD measurements as well as theoretical calculations find that the pressurized $Ba_{0.61}K_{0.39}Mn_2Bi_2$ is a robust AFM metal. No evidence for superconductivity is found under pressure up to 35.6 GPa, however, a pressure-induced structural transition from T phase to OR phase is observed at ~20 GPa. Theoretical calculations demonstrate that the values of magnetic moment on Mn in the T phase and OR phase are nearly identical at ~3.4 $\mu_B$, suggesting that the K-doped $BaMn_2Bi_2$ is a strong Hund's AFM metal with a hybridization of localized spin electrons and itinerant electrons. The robust AFM order in $Ba_{0.61}K_{0.39}Mn_2Bi_2$ essentially prevents the emergence of superconductivity.

**Acknowledgements**

This work in China is supported by the NSF of China (Grant No. 91321207,1190024, 11334012), 973 projects (Grant No.2011CBA00100, 2012CB821400 and 2010CB923000) and Chinese Academy of Sciences. The work in the USA has been supported by the U.S. Department of Energy, Basic Energy Sciences, Materials Sciences and Engineering Division.

†Correspondence to llsun@iphy.ac.cn and zhxzhao@iphy.ac.cn .

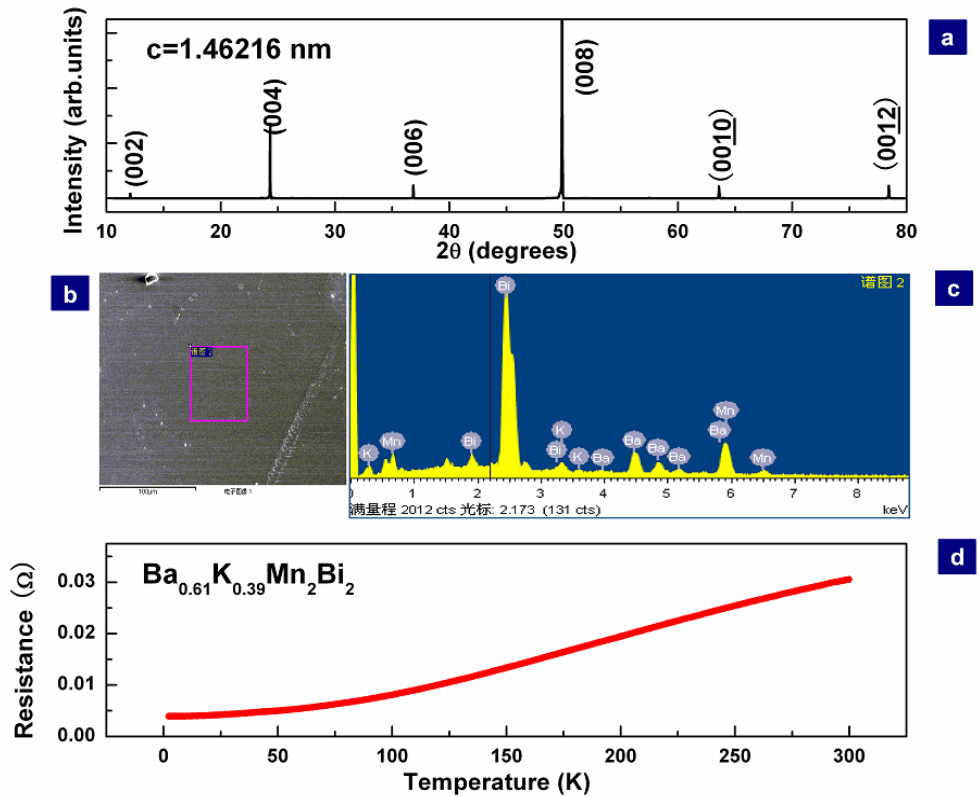

Fig. 1 (a) X-ray diffraction pattern on $Ba_{0.61}K_{0.39}Mn_2Bi_2$ single crystal collected at ambient pressure by using a Rigaku Ultima IV diffractometer. (b) Image of the investigated single crystal, taken by a scanning electronic microscopy. (c) Representative energy dispersive X-ray analysis results measured on the squared area of (b), giving an average actual composition of $Ba_{0.61}K_{0.39}Mn_2Bi_2$. (d) Resistance-temperature curve for $Ba_{0.61}K_{0.39}Mn_2Bi_2$ at ambient pressure.

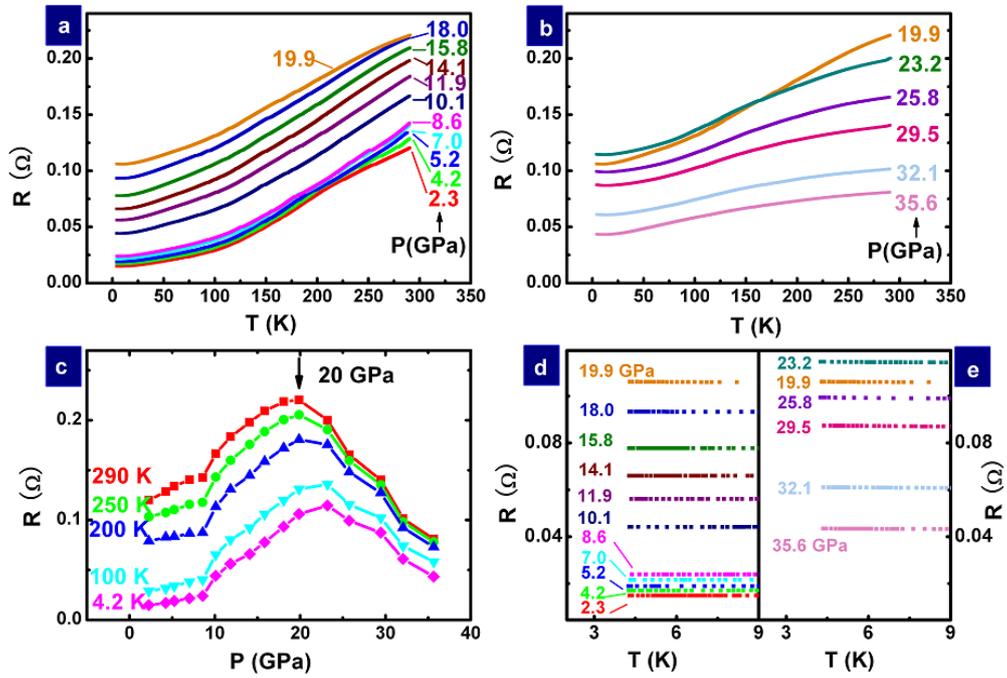

Fig. 2 (a) and (b) Temperature dependent resistance of $Ba_{0.61}K_{0.39}Mn_2Bi_2$ at different pressures before and after the T-to-OR phase transition. (c) Pressure dependence of resistance measured at different temperatures, displaying a dome feature centered at 20-23 GPa. (d) and (e) The zoomed *R-T* curves at different pressures, showing no evidence of superconductivity.

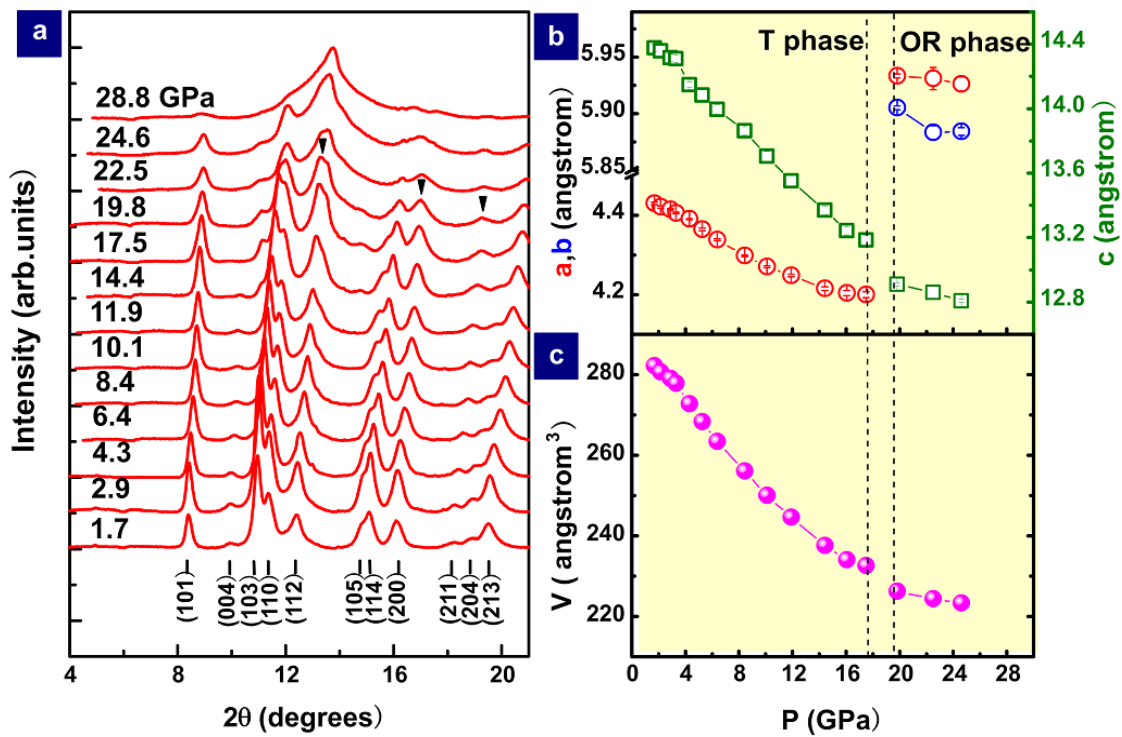

Fig. 3(a) Representative XRD patterns for $Ba_{0.61}K_{0.39}Mn_2Bi_2$ at various pressures. (b) Pressure dependence of lattice constant *a* (red circle, $a=b$ in the T phase), *b* (blue circle) and *c* (green square). (c) Volume as a function of pressure in the tetragonal (T) and orthorhombic (OR) phases.

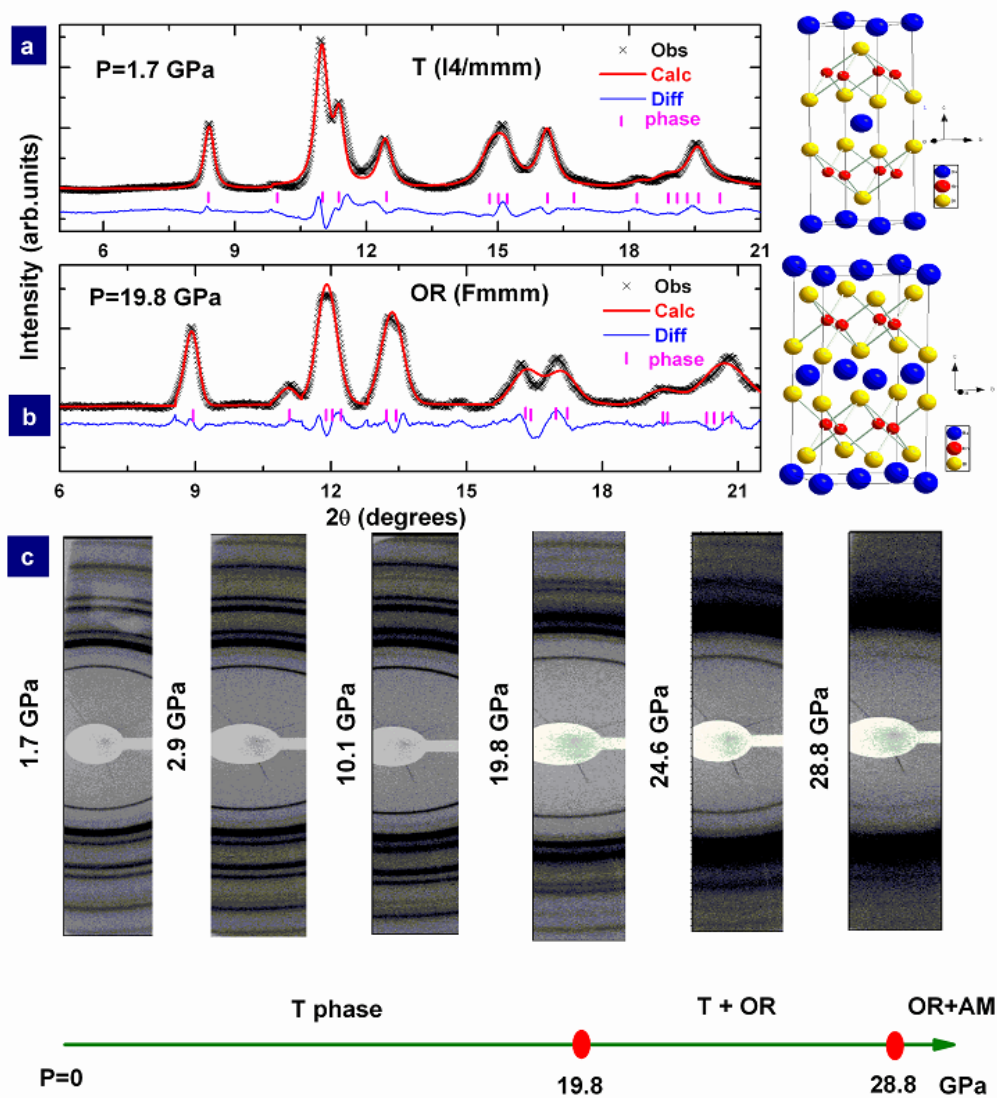

Fig. 4 (a) and (b) Rietveld refinement results of the X-ray diffraction patterns at 1.7 GPa and 19.8 GPa in the tetragonal (T) phase (*I*4/*mmm*) and the orthorhombic (OR) phase (*Fmmm*) with the corresponding crystal structures shown on the right. (c) Representative X-ray diffraction images under pressure, showing structure evolution with pressure.

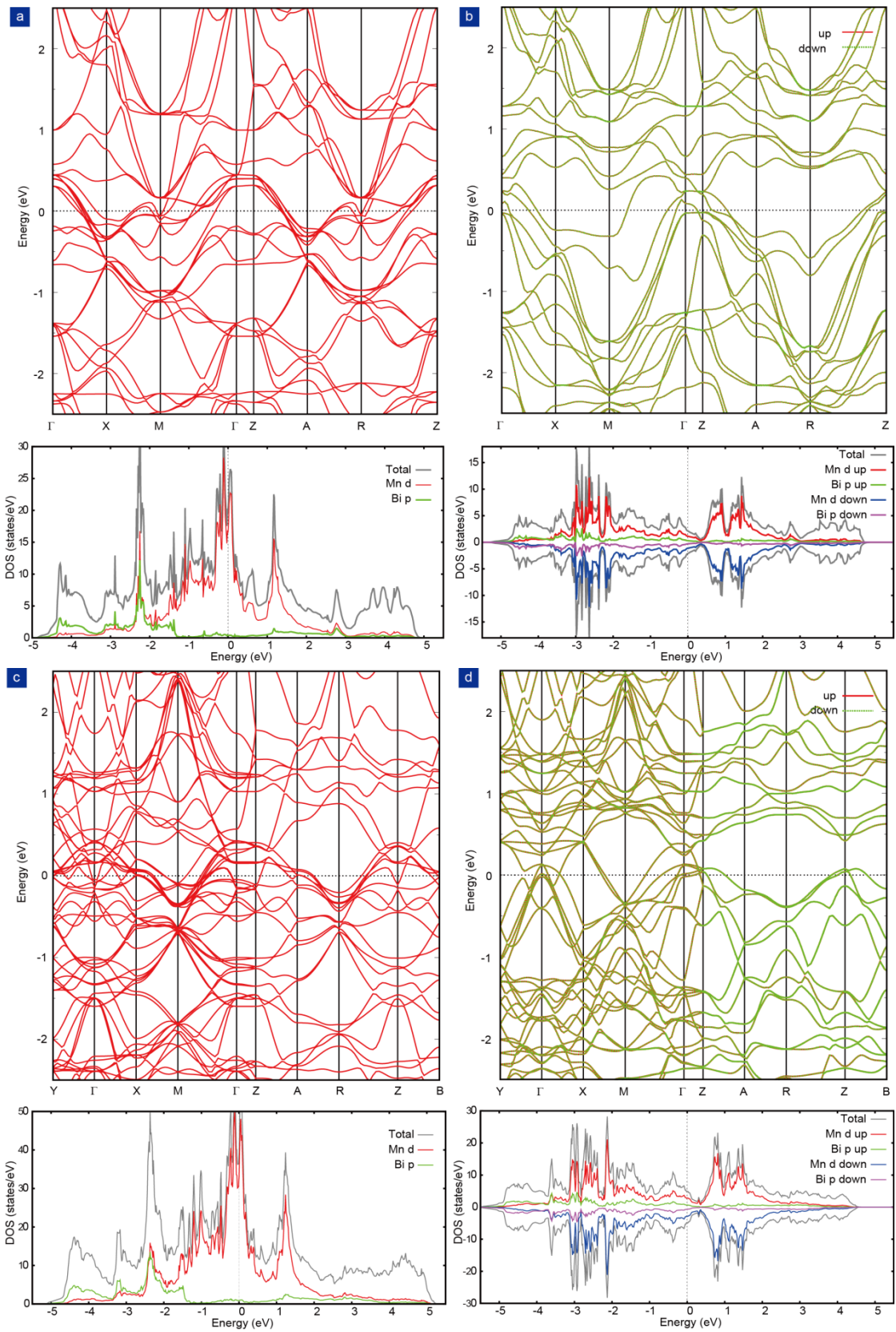

Fig. 5 (a) and (b) Band structures and density-of-states for the tetragonal phase of $Ba_{0.61}K_{0.39}Mn_2Bi_2$ at 17.49 GPa in the paramagnetic and antiferromagnetic states, respectively. (c), (d) Band structures and density-of-states of the orthorhombic phase of $Ba_{0.61}K_{0.39}Mn_2Bi_2$ at 24.6 GPa in the paramagnetic and antiferromagnetic states, respectively.